\documentclass[prl,twocolumn,nofootinbib,reprint]{revtex4-1}
\usepackage{amsmath,amssymb,amsthm,bm,braket,ascmac,bbm}
\usepackage{graphicx}
\usepackage{color} 
\usepackage{hyperref}

\theoremstyle{definition}

\newcommand{\MeV}{\  {\rm MeV} }

\newcommand{\lmk}{\left(}  
\newcommand{\rmk}{\right)}

\newcommand{\eq}[1]{Eq.~(\ref{#1})}
\newcommand{\Mpl}{M_{p}}

\begin{document}
\title{
Gravitational Waves and Dark Radiation from Dark Phase Transition: 
\\
Connecting NANOGrav Pulsar Timing Data and Hubble Tension
}
\author{ Yuichiro Nakai$^1$, Motoo Suzuki$^1$, Fuminobu Takahashi$^{2,3}$, and Masaki Yamada$^{2,4}$}
\affiliation{\vspace{2mm} \\
$^1$Tsung-Dao Lee Institute and School of Physics and Astronomy, \\Shanghai
Jiao Tong University, 800 Dongchuan Road, Shanghai, 200240 China \\
$^2$
Department of Physics, Tohoku University, Sendai, Miyagi 980-8578, Japan \\
$^3$
Kavli IPMU (WPI), UTIAS, The University of Tokyo, 
Kashiwa, Chiba 277-8583, Japan \\
$^4$
Frontier Research Institute for Interdisciplinary Sciences, Tohoku University, 
Sendai, Miyagi 980-8578, Japan
}

\begin{abstract}

 Recent pulsar timing data reported by the NANOGrav collaboration may indicate the existence of a stochastic gravitational wave background around $f \sim 10^{-8}$ Hz. We explore a possibility to generate such low-frequency gravitational waves from a dark sector phase transition. Assuming that the dark sector is completely decoupled from the visible sector except via the gravitational interaction, we find that some amount of dark radiation should remain until present. The NANOGrav data implies that the amount of dark radiation is close to the current upper bound, which may help mitigate the so-called Hubble tension. If the existence of dark radiation is not confirmed in the future CMB-S4 experiment, it would imply the existence of new particles feebly interacting with the standard model sector at an energy scale of ${\cal O}(1$\,-\,$100)$\,MeV.
 \end{abstract}

\begin{flushright}
TU-1109~
IPMU20-0100
\end{flushright}

\maketitle

%#######################
{\bf Introduction.--}
The direct detection of gravitational waves (GWs) by the LIGO and Virgo collaborations
\cite{Abbott:2016blz}
has opened up a fascinating era of astronomy and cosmology that looks at our Universe through entirely new eyes.
Since GWs propagate without interaction, their detection enables us to probe physics in the early Universe.
While ground-based interferometers such as LIGO and Virgo have the best sensitivity
at frequencies of around 100 Hz,
searches for GWs with lower frequencies have been also conducted or planed
in various types of experiments.
Future space-based GW observers such as LISA~\cite{Seoane:2013qna}, DECIGO~\cite{Seto:2001qf}, and BBO~\cite{Crowder:2005nr,Corbin:2005ny} have
the best sensitivity at frequencies of order mHz.
GWs with even lower frequencies of $\mathcal{O}(10^{-9}) \, \rm Hz$
are searched for by pulsar timing array (PTA) experiments
such as EPTA~\cite{Lentati:2015qwp}, PPTA~\cite{Manchester_2013}, and NANOGrav~\cite{McLaughlin:2013ira,Brazier:2019mmu}.
Now, the discovery of such low frequency GWs may be right around the corner. 

Recently, the NANOGrav collaboration of a PTA experiment
has analyzed their 12.5 years of data and
reported a signal that may be interpreted as a GW background
\cite{Arzoumanian:2020vkk}.
One possible source of the signal is an astrophysical GW background
generated by mergers of super-massive black-hole binaries
\cite{Rajagopal:1994zj,Phinney:2001di,Jaffe:2002rt,Wyithe:2002ep}.
Another possibility is a stochastic GW background
emitted by a cosmic-string network in the early Universe,
which can give a favored flat spectrum of frequencies in the GW energy density
\cite{Ellis:2020ena,Blasi:2020mfx} (see also, e.g.,  Refs.~\cite{Vilenkin:1981bx,Vachaspati:1984gt,Ringeval:2005kr,Siemens:2006yp,Kawasaki:2010yi,BlancoPillado:2011dq,Blanco-Pillado:2017rnf,Ringeval:2017eww,King:2020hyd} for earlier works).
A stochastic GW signal associated to primordial black hole formation 
has also been investigated \cite{Vaskonen:2020lbd,DeLuca:2020agl}.

In this paper, we explore an interpretation of the reported NANOGrav signal
in terms of a stochastic GW background generated by a new strongly first-order phase transition
which occurred in a dark sector.
Generation of GWs from a strongly first-order phase transition has been actively discussed.
A possible detection of GWs from the QCD phase transition through PTAs was 
pointed out in Ref.~\cite{Caprini:2010xv}.
GWs from a supercooled electroweak phase transition
and their detection with PTAs were discussed in Ref.~\cite{Kobakhidze:2017mru} (see also Ref.~\cite{Iso:2017uuu}).
Such a supercooled phase transition has been known to be realized in warped extra dimension models or
their holographic duals~\cite{Creminelli:2001th,Randall:2006py,vonHarling:2017yew,Megias:2018sxv,Baratella:2018pxi,Agashe:2019lhy,Fujikura:2019oyi,vonHarling:2019gme,DelleRose:2019pgi}.
GWs from dark sector phase transitions were explored in Refs.~\cite{Schwaller:2015tja,Jaeckel:2016jlh,Croon:2018erz,Fujikura:2018duw,Bhoonah:2020oov} 
though they considered the case in which the dark sector is not completely decoupled from the visible sector. 
In particular, Ref.~\cite{Schwaller:2015tja} discussed a range of GW frequencies covered by PTAs. In this paper, we focus on the first order phase transition in the dark sector decoupled from the visible sector.
The production of GW in such a decoupled dark sector was studied in Refs.~\cite{Breitbach:2018ddu,Fairbairn:2019xog}. As we shall see below, however, the estimated GW amplitude in Ref.~\cite{Breitbach:2018ddu} was underestimated, and we will also use the updated GW spectrum taking account of the effective lifetime of the GW sources.
We derive a consistency relation between the GW energy density to explain the reported NANOGrav signal
and the allowed abundance of dark radiation components in a dark sector.
If the NANOGrav signal is confirmed, it predicts
a future discovery of dark radiation components in our Universe. 
The amount of dark radiation will provide us with information on details of the phase transition.
Interestingly, in some (rather realistic) parameter space 
our scenario predicts $\Delta N_{\rm eff} \sim 0.4$\,-\,$0.5$ which can ameliorate the so-called Hubble tension~\cite{Bernal:2016gxb,Aghanim:2018eyx,Blinov:2020hmc}.

The rest of the paper is organized as follows.
In the next section,
we derive the relation between the amount of dark radiation and the amplitude of stochastic GWs generated 
by a first-order phase transition in a dark sector.
Then we compare the NANOGrav data with the GWs generated by the phase transition 
and discuss its implications for the amount of dark radiation. 
Finally, we conclude and discuss a possible model of the dark sector. 
In Appendix, 
we quote and summarize the GW spectrum generated from the first order phase transition 
that is used in our numerical calculations.

%#######################
\vspace{0.1cm}
{\bf GW and dark radiation.--}
We consider the case in which 
the GW signal reported by NANOGrav comes from the first-order phase transition in the dark sector. 
As we stated in the introduction, we assume that the dark sector is coupled to the visible sector only via the gravitational interaction and study its implications for dark radiation. 
We do not specify how the first-order phase transition occurs in the dark sector,
but describe nature of the phase transition in terms of phenomenological parameters such as the wall velocity $v_w$, 
duration of the phase transition $\beta^{-1}$, 
and an efficiency factor $\kappa_i$ defined later and/or in Appendix.

As the dark sector is decoupled from the visible sector, it is reasonable to assume
that the entropy is separately conserved in each sector until the beginning of the phase transition and after the end of the phase transition. 
We denote the entropy ratio between the two sectors before the phase transition as $R_i$: 
\begin{align}
 R_i \equiv 
 \left. 
 \frac{s_{\rm D}}{s_{\rm vis}} 
 \right\vert_{T > T_*}, 
\end{align}
where $s_D$ ($s_{\rm vis}$) is the entropy density in the dark (visible) sector 
and 
$T_*$ is the temperature of the visible sector at the time of the first order phase transition. 
As we will see, we are interested in the case in which the dark sector never dominates the energy density of the Universe, so that the entropy production does not dilute particles in the visible sector much, including the baryon asymmetry. 
The ratio $R_i$ is considered to be determined by the reheating process after inflation. 
We take it as a free parameter, but expect 
it to be neither very small nor very large, 
as in the case of the universal reheating.
When the phase transition is a strong first-order phase transition, a large entropy is generated from the latent heat in the dark sector. Denoting the entropy production factor by $\Delta (\geq 1)$, 
we can express the entropy ratio after the phase transition as 
\begin{align}
 R = \Delta \, R_i. 
\end{align}
If one assumes 
that the phase transition is not a strongly supercooled one, $R\sim R_i$ 
and is neither very small nor very large.

After the phase transition, some amount of energy (and entropy) should remain in the dark sector that is decoupled from the visible sector.
Then, the lightest particle(s) in the dark sector should be (almost) massless and behave as dark radiation, 
since otherwise the abundance of the remnant in the dark sector would easily exceed the observed dark matter density and overclose the Universe.\footnote{
It is possible that dark matter is explained by a fraction of the remnants in the
dark sector, but this does not change the following discussion.
} 
The amount of dark radiation at the recombination epoch, $\rho_{\rm DR,0}$, is calculated as 
\begin{align}
 \rho_{\rm DR,0} = \rho_{\rm rad,0} R^{4/3}
 \lmk \frac{g_{*0}^{\rm (D)}}{g_{*0}} \rmk 
 \lmk \frac{g_{*s0}}{g_{*s0}^{\rm (D)}} \rmk^{4/3},
\end{align}
where $g_{*0}$ ($g_{*s0}$) and $g_{*0}^{\rm (D)}$ ($g_{*s0}^{\rm (D)}$) are the effective numbers of relativistic degrees of freedom for the energy (entropy) densities in the visible sector and the dark sector, respectively, 
and $\rho_{\rm rad,0}$ is the energy density of photons and three neutrinos in the visible sector. 
The subscript $0$ represents the value at the recombination epoch. 
This gives the extra effective neutrino number of 
\begin{align}
 \Delta N_{\rm eff} \simeq 
 0.49 \times \lmk \frac{R}{0.13} \rmk^{4/3}
 \lmk \frac{g_{*0}^{\rm (D)}}{g_{*0}} \rmk 
 \lmk \frac{g_{*s0}}{g_{*s0}^{\rm (D)}} \rmk^{4/3}.
 \label{Delta N_eff}
\end{align}
The Planck data combined with the BAO observation and the local Hubble measurement gives the constraint~\cite{Riess_2018, Aghanim:2018eyx}
\begin{align}
 N_{\rm eff} = 3.27 \pm 0.15 \quad (68\% \, {\rm C.L.}).
 \label{N_eff}
\end{align}
The prediction in the standard cosmology is $N_{\rm eff}^{\rm (std)} = 3.046$. 
Note that there is a tension between the local measurement of the Hubble parameter and the Hubble parameter inferred by the Planck and BAO with $\Delta N_{\rm eff} = 0$. 
The tension is known to be relaxed if $\Delta N_{\rm eff} \simeq 0.4$\,-\,$0.5$~\cite{Bernal:2016gxb,Aghanim:2018eyx,Blinov:2020hmc},
which roughly amounts to $R \simeq 0.1$\,-\,$0.2$.

Now we shall relate the entropy ratio and the density parameter of the GW. 
We define 
\begin{align} 
\label{eq:alpha}
 \alpha' \equiv \frac{\rho_{\rm vac}}{\rho_{\rm rad,tot} (T_*)}, 
\end{align}
where $\rho_{\rm vac}$ is the false vacuum energy in the dark sector at the phase transition 
and $\rho_{\rm rad,tot}(T_*)$ ($= \rho_{\rm rad} (T_*) + \rho_{\rm DR} (T_{*i}^{\rm (D)}) \simeq 3 H_*^2 \Mpl^2$) 
is the total radiation energy density just before the phase transition. 
Here, we denote $T_{*i}^{\rm (D)}$ and $T_{*f}^{\rm (D)}$ as the temperatures of the dark radiation 
before and after the phase transition, respectively. 
After the phase transition, the vacuum energy is converted into the radiation energy in the dark sector. 
Using the entropy production $\Delta$, we can relate $\alpha'$ and $R$ as
\begin{align}
 &\alpha' = \lmk \frac{r}{1+r} \rmk
 \lmk \frac{\alpha}{1+\alpha} \rmk, 
  \label{alpha'}
 \\
 &r \equiv \frac{\rho_{\rm DR}}{\rho_{\rm rad}} 
 = \lmk \frac{g_*^{\rm (D)} (T_{*f}^{\rm (D)})}{g_*(T_*)}\rmk 
 \lmk \frac{g_{*s}(T_*)}{g_{*s}^{\rm (D)}(T_{*f}^{\rm (D)})} \rmk^{4/3} R^{4/3}, 
 \label{r}
 \\
 &\alpha \equiv \frac{\rho_{\rm vac}}{\rho_{\rm DR}(T_{*i}^{\rm (D)})} 
 \nonumber\\
 &~~= \Delta^{4/3} \lmk \frac{g_{*}^{\rm (D)}(T_{*f}^{\rm (D)})}{g_{*}^{\rm (D)}(T_{*i}^{\rm (D)})} \rmk \lmk \frac{g_{*s}^{\rm (D)}(T_{*i}^{\rm (D)})}{g_{*s}^{\rm (D)}(T_{*f}^{\rm (D)})} \rmk^{4/3} - 1, 
\end{align}
where we assume the instantaneous reheating after the phase transition and use $\rho_{\rm vac} + \rho_{\rm DR}(T_{*i}^{\rm (D)}) = g_*^{\rm (D)} (\pi^2/30) (T_{*f}^{\rm (D)})^4$.

We are interested in the case in which 
the GW is efficiently emitted. 
From this consideration, 
we assume that the phase transition 
is a strong first-order phase transition, 
which implies $\Delta \gtrsim {\cal O}(1)$. 
In this case, 
we can neglect the second parenthesis in \eq{alpha'}. 
We also note that $R \lesssim 0.1$, which leads to 
$\alpha' \simeq r \ll 1$. 
Combining Eqs.~(\ref{Delta N_eff}) and (\ref{alpha'}), 
we obtain 
\begin{align}
 \alpha' \simeq 
 0.07 \lmk \frac{\Delta N_{\rm eff}}{0.5} \rmk
 \lmk \frac{g_{*0}}{g_*} \rmk
 \lmk \frac{g_{*s}}{g_{*s0}} \rmk^{4/3}
 \lmk \frac{g_{*}^{\rm (D)}}{g_{*0}^{\rm (D)}} \rmk
 \lmk \frac{g_{*s0}^{\rm (D)}}{g_{*s}^{\rm (D)}} \rmk^{4/3},
 \nonumber\\
 \label{alpha'2}
\end{align}
where $g_*$'s are evaluated at $T = T_*$ or $T_{*f}^{\rm (D)}$ like \eq{r}.

We also define $\kappa_i$ that generically represents 
the fraction of latent heat converted to the GW source labeled by $i$, 
which we specify later and in Appendix. 
Since the GW comes from the quadrupole moment, 
the resulting GW amplitude is proportional to the energy density of the source squared. 
We also note that the amplitude of stochastic GW is involved with two time integrals. 
Motivated by this observation and taking into account the redshift factor, 
we factorize the density parameter of the GW at present as (see, e.g., Refs.~\cite{Kamionkowski:1993fg,Caprini:2009fx,Jinno:2016vai})
\begin{align}
 \Omega_{\rm GW,0} &= \sum_i \Omega_{\rm rad,0} \lmk \frac{g_{*}(T_*)}{g_{*0}} \rmk
 \lmk \frac{g_{*s0}}{g_{*s}(T_*)} \rmk^{4/3}
 \nonumber\\
 &~~~~~~~~ \times \lmk \frac{H_*}{\beta} \rmk^2
 \lmk \frac{\kappa_i \alpha'}{1+\alpha'} \rmk^2 
  \tilde{\Omega}_{{\rm GW},i}, 
  \label{Omega_gw}
\end{align}
where $H_* \equiv H(T_*)$ and $\Omega_{\rm rad,0} h^2= 4.16 \times 10^{-5}$~\cite{Fixsen:2009ug} 
with $h$ being the reduced Hubble parameter.
Here, $\beta^{-1}$ represents the duration of the phase transition, defined by 
\begin{align}
 \beta = \frac{1}{\Gamma} \frac{d \Gamma}{dt},
\end{align}
with $\Gamma$ being the bubble nucleation rate. 
The remaining factor $\tilde{\Omega}_{{\rm GW},i}$ is determined by numerical simulations and/or (semi-)analytic calculations (see Appendix). 
In particular, 
it depends only on $k/\beta$ and the bubble wall velocity $v_w$ 
for the GW emission from the bubble collision under a certain assumption~\cite{Jinno:2016vai}.

Using \eq{alpha'2}, 
we can see that the density parameter of the GW is proportional to $\Delta N_{\rm eff}^2$. 
These are the formula 
that relates the amplitude of the GW to
the amount of dark radiation $\Delta N_{\rm eff}$. 
This can be applied to any models where the GWs are emitted in a dark sector that is completely decoupled from the visible sector except via the gravitational interaction. 
Contrary to the ordinary scenario where the phase transition occurs in the visible sector, 
the amplitude of the GW is determined by $\alpha'$ rather than $\alpha$. Here
$\alpha'$ cannot be much larger than the order $0.1$ 
due to the constraint on the dark radiation (see \eq{N_eff}).

%#######################
\vspace{0.1cm}
{\bf GW and NANOGrav Pulsar Timing Data.--}
There are three possible sources of the GWs from the first-order phase transition; 
the collision of true-vacuum bubbles~\cite{Turner:1990rc,Kosowsky:1991ua,Kosowsky:1992vn,Turner:1992tz,Kamionkowski:1993fg,Caprini:2007xq,Huber:2008hg,Jinno:2016vai,Jinno:2017fby,Konstandin:2017sat,Cutting:2018tjt,Cutting:2020nla,Lewicki:2020jiv,Ellis:2020nnr}, the sound waves~\cite{Hindmarsh:2013xza,Giblin:2014qia,Hindmarsh:2015qta,Hindmarsh:2017gnf}, and the turbulence~\cite{Kamionkowski:1993fg,Caprini:2006jb,Caprini:2009yp,Kosowsky:2001xp,Gogoberidze:2007an,Niksa:2018ofa}. 
For each source, we define $\kappa_i$ with $i = {\rm bubble, \ SW, \ turb}$. We summarize the GW spectra produced from these sources in Appendix.

Note that $\kappa_i$ are determined by the dynamics in the dark sector, so that they are related to parameters in the dark sector such as $\alpha$ rather than $\alpha'$.%
\footnote{
In Ref.~\cite{Breitbach:2018ddu}, 
they assumed that $\kappa_i$ are determined by $\alpha'$ (corresponding to their $\alpha$), which led them to use smaller $\kappa_i$ than the actual values.
As a result, their estimated GW was underestimated compared to ours.
}
Since $\alpha$ can be larger than of order unity, 
some $\kappa_i$ can be as large as of order unity~\cite{Kamionkowski:1993fg,Espinosa:2010hh}. 
To compare the NANOGrav data, 
we provide the GW spectra for two cases using the formula written in Appendix. 
The first case is the GW spectrum only from the sound wave and turbulence 
while the second case is the one from all three sources. 
We take $\kappa_{\rm turb} = 0.1 \kappa_{\rm SW}$ for both cases, 
as suggested by numerical simulations~\cite{Caprini:2015zlo}. 
We assume the maximal efficiency, 
such as $\kappa_{\rm SW} = 1/1.1$ in the former case 
and $\kappa_{\rm bubble} = \kappa_{\rm SW} =1/2.1$ for the latter case. 

In our numerical calculations, we take $v_w \simeq 1$ and $(\alpha / (1 + \alpha)) \simeq 1$ and set $g_*^{\rm (D)}=g_{*0}^{\rm (D)}$ and $g_{*s}^{\rm (D)}=g_{*s0}^{\rm (D)}$ for simplicity. 
The duration of the phase transition, $\beta/H_*$, 
is typically about $100$ for $T_* \sim 1-100 \MeV$ 
though it can be as small as of order unity depending on models~\cite{Hogan:1984hx}. 
We take it to be a free parameter within $(1$\,-\,$100)$ 
to show examples. 

%%%%
\begin{figure}
\centering
  \includegraphics[width=0.9\linewidth]{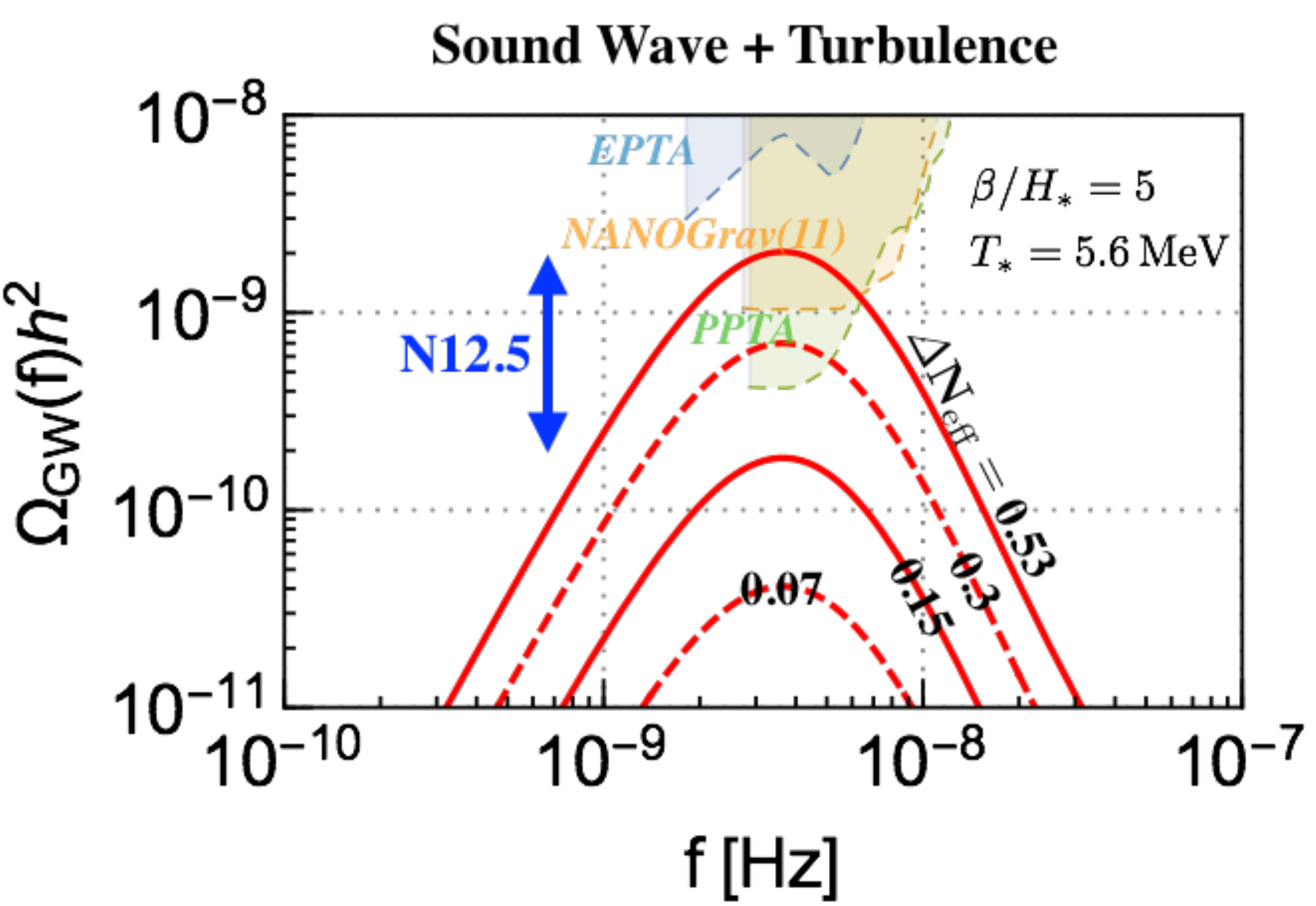}
   \\
  \vspace{0.5cm}
  \includegraphics[width=0.9\linewidth]{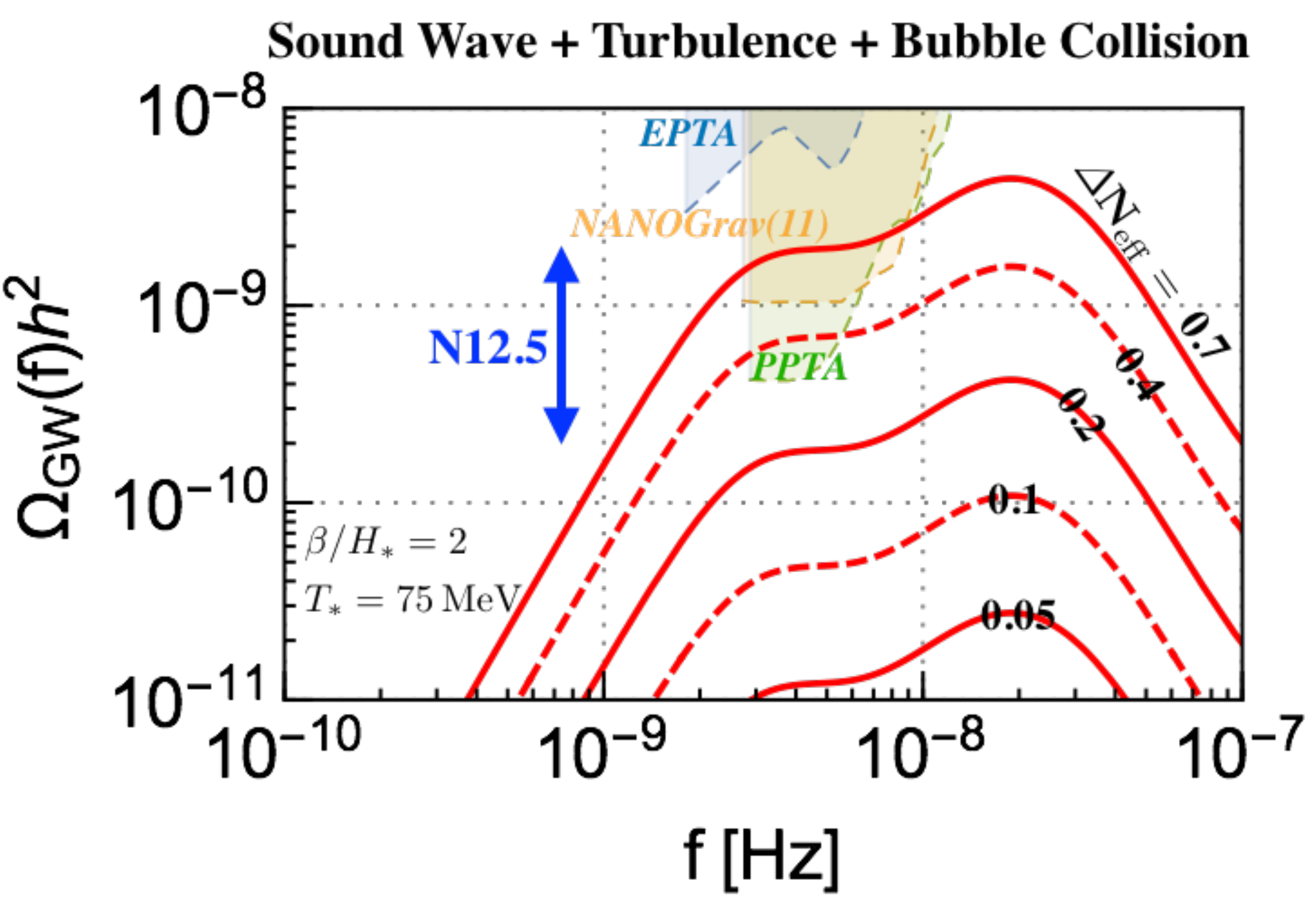}
   \caption{
   GW spectra produced from the first-order phase transition in the dark sector 
   with (lower figure) and without (upper figure) the contribution from the bubble collision. 
   The GW amplitude is related to the amount of the dark radiation, which is a remnant in the dark sector. The vertical arrow (blue) represents the range of the amplitude favored by the NANOGrav $12.5$ year pulsar timing data. 
   }
\label{fig:spectrum}
\end{figure}
%%%%  

%%%%
\begin{figure}
\centering
  \includegraphics[width=0.9\linewidth]{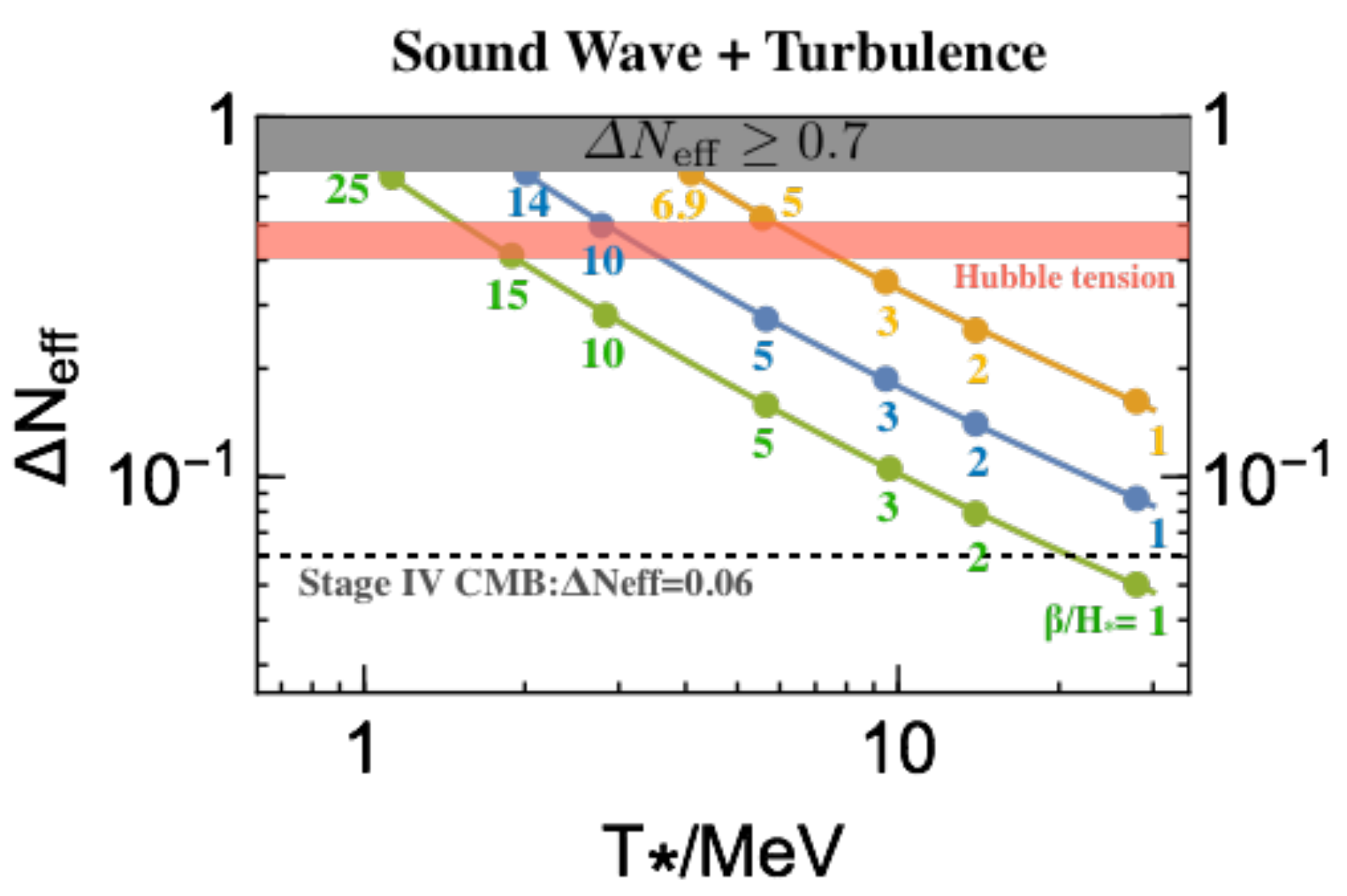}
  \\
  \vspace{0.5cm}
  \includegraphics[width=0.9\linewidth]{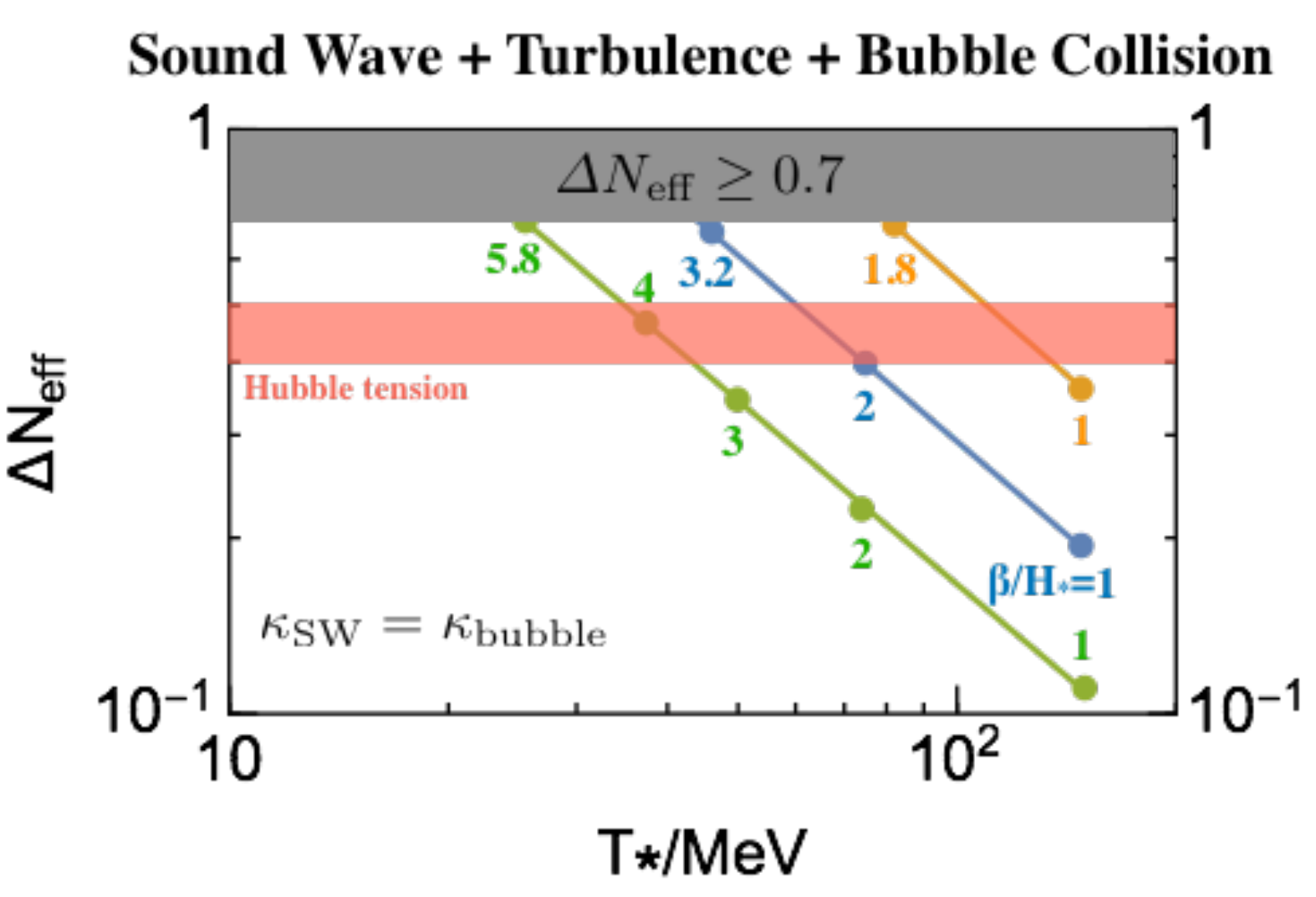}
   \caption{The parameter region favored by the NANOGrav data on the $\Delta N_{\rm eff}$\,-\,$T_*$ plane. 
   The yellow (top), blue (middle), and green (bottom) lines correspond to the case with the peak amplitude of $\Omega_{\rm GW}(f)h^2=2\times 10^{-9}\ ,6\times10^{-10}\ ,2\times 10^{-10}$ at the peak frequency of $f=3.7\times 10^{-9}$\,Hz, respectively. The numbers shown near the dots on each line 
   represent the corresponding $\beta/H_*$. 
   The gray-shaded region is excluded by Planck and BAO while the red-shaded regions are favored to ameliorate the Hubble tension. 
   }
\label{fig:result}
\end{figure}
%%%%  

Figure~\ref{fig:spectrum} shows the GW spectra produced by 
the first-order phase transition 
and the NANOGrav 12.5 year results with older experimental constraints. 
In the upper figure, 
we show the GW signals from the sound wave and turbulence with
$\beta/H_*=5$ and $T_*=5.6\,{\rm MeV}$. 
In the lower figure, 
we include all three sources 
with $\beta/H_*=2$ and $T_*=75\,{\rm MeV}$.
The red lines represent the GW spectrum 
for several values of $\Delta N_{\rm eff}$. 
The double-headed arrow shows the amplitude favored by the NANOGrav 12.5 year signal within $2$-$\sigma$ posterior contour when the spectrum is assumed to be flat~\cite{Arzoumanian:2020vkk}.
The blue, yellow and green shaded regions are the previous bounds from EPTA~\cite{Lentati:2015qwp}, NANOGrav$(11{\rm yr})$~\cite{Arzoumanian:2018saf}, and PPTA~\cite{Shannon:2015ect}.
While the results of the NANOGrav 12.5yr are not compatible with the older constraints, the tension is understood by the improvement of the pulsar red noise treatment in the NANOGrav 12.5yr~\cite{Arzoumanian:2020vkk}.

In Ref.~\cite{Arzoumanian:2020vkk}, 
they fitted 
the NANOGrav pulsar timing data by a power law spectrum and a broken power law spectrum 
and concluded that the data favor spectrum with a power of $-0.5$\,-\,$1.5$ 
around $f = 3.7\times 10^{-9}$. 
However, the most relevant data are two bins at the first and second lowest frequencies, which can be fitted by a spectrum with a peak at $f \simeq 3.7\times 10^{-9}$, just like the ones in the upper figure in Fig.~\ref{fig:spectrum}. 
The spectrum shown in the lower figure 
has a slightly positive power in the relevant frequency scale, 
which is also favored by the NANOGrav data.

We also plot $\beta/H_*$ on the $T_*$\,-\,$\mit\Delta N_{\rm eff}$ plane in Fig.~\ref{fig:result}. 
The upper figure corresponds to the case for the GW produced by the sound wave and turbulence 
while the lower one corresponds to the case for all three sources. 
The yellow, blue, and green lines correspond to the case with the peak amplitude of $\Omega_{\rm GW}(f)h^2=2\times 10^{-9}\ ,6\times10^{-10}\ ,2\times 10^{-10}$ for the peak frequency of $f=3.7\times 10^{-9}$\,Hz, respectively, 
which are implied by the NANOGrav data. 
In more details, in the lower figure, we require the peak of the bubble collision corresponds to the NANOGrav data implied values. 
However, we note that the NANOGrav data can also be
explained by the highest peak of the spectrum that comes from the sound wave (rather than the one comes from the bubble collision). In this case, 
the results are almost the same as the one in the upper figure even if we include all three sources.

The numbers shown near the dots on the lines represent $\beta/H_*$. 
We note that $\beta / H_*$ is typically as large as about $100$ for realistic cases but can be as small as ${\cal O}(1)$ for the case of, e.g., supercooling phase transition. 
The gray-shaded region is excluded by the Planck and BAO observations while the red-shaded region is favored to ameliorate the Hubble tension. 
The black-dotted line in the upper figure is the 
prospected $2\sigma$ 
sensitivity of the stage-IV ground-based detector, CMB-S4~\cite{Abazajian:2016yjj}. 
Our scenario predicts $\Delta N_{\rm eff}$ that can be measured in the near future, by explaining the NANOGrav data by the phase transition in the dark sector. In fact, some parameter space is also favored to ameliorate the Hubble tension.

\vspace{0.1cm}
{\bf Discussions and conclusions.--}
We have discussed the implications of NANOGrav data for the case in which 
the stochastic GWs are produced from the first order phase transition in the dark sector, 
assuming that the dark sector is completely decoupled from the visible sector except via the gravitational interaction. 
Since there must be a remnant in the dark sector after the phase transition, 
the GW amplitude is related to the amount of dark radiation. Interestingly,
the predicted abundance of dark radiation is within the reach of CMB-S4 in the near future.
In some parameter space, 
the amount of dark radiation is as large as the one preferred to ameliorate the Hubble tension. 
In other words, the signal observed by NANOGrav is closely related to the Hubble tension.

If the dark radiation is not observed in the near future, 
it may imply that the dark sector is coupled to the visible sector. In this case, $T_*$ should be larger than of order $1$\,-\,$10 \MeV$ so that the phase transition and the subsequent energy injection to the visible sector should not spoil the success of the Big Bang nucleosynthesis. Note that the Big Bang nucleosynthesis bound is significantly weaker in the case of the dark sector completely decoupled from the visible sector.

The peak frequency of the GW is related to 
the energy scale of the phase transition, which we found is $T_* = {\cal O}(1$\,-\,$100) \MeV$. 
As the energy scale is close to the QCD scale, 
one may think of a possibility that the GW signal comes from the phase transition of a dark QCD. 
This is the case, e.g., of a parallel world, 
where 
the dark sector has a similar (but not exactly the same) structure to the visible sector~\cite{Higaki:2013vuv}. 
If the parameters in the dark sector are slightly different from the ones in the visible sector, 
it is possible that the dark-QCD phase transition is the strong first-order phase transition 
and its energy scale is $T_* = {\cal O}(1$\,-\,$100) \MeV$. 
This scenario also provides a dark-matter candidate, which is the dark neutron. 
Since the dark neutron can have a sizable self-interaction cross section, 
this scenario might be confirmed by astrophysical observations for the dark-matter density profile. 
Also the dark neutrinos behave as hot dark matter, which will mitigate the $\sigma_8$ tension
when dark radiation is introduced to solve the Hubble tension.

So far we have focused on the GWs generated from a first order phase transition in 
the dark sector, but it is also possible to generate  similar GW spectra by 
considering decays of topological defects such as domain walls and/or cosmic strings (see e.g. Ref.~\cite{Nakayama:2016gxi,Saikawa:2017hiv}).
Our consistency relation can easily be extended to these cases as well.
Also, the clockwork QCD axion~\cite{Higaki:2015jag} is known to have an extremely complicated 
network of strings and domain walls which store a large amount of energy~\cite{Higaki:2016jjh,Long:2018nsl}. The string/wall network
annihilate around the QCD phase transition, which may produce sizable
GWs with frequencies covered by the PTA experiments~\cite{Higaki:2016jjh}.

%---------------SECTION------------------%
%
\section*{Acknowledgements}
We thank Kohei Fujikura, Marek Lewicki, and Graham White for pointing out the suppression factor for the sound-wave period. 
Y.N. would like to thank Kohei Fujikura and Keisuke Harigaya for discussions. Y.N. is grateful to Kavli IPMU for their hospitality during the COVID-19 pandemic.
M.S. would like to thank Ryo Namba for discussions. 
F.T. was supported by JSPS KAKENHI Grant Numbers
17H02878, 20H01894 and by World Premier International Research Center Initiative (WPI Initiative), MEXT, Japan. 
M.Y. was supported by Leading Initiative for Excellent Young Researchers, MEXT, Japan. 
%
%---------------SECTION------------------%

\vspace{0.5cm}

{\bf Appendix: GW spectrum.--}
In this appendix, we quote and summarize explicit values of 
$\tilde{\Omega}_{\rm GW}$ (see, e.g., Refs.~\cite{Caprini:2015zlo,Caprini:2019egz} for more detailed discussion). 

There are three possible sources of the GWs from the first-order phase transition; 
the collision of true-vacuum bubbles~\cite{Turner:1990rc,Kosowsky:1991ua,Kosowsky:1992vn,Turner:1992tz,Kamionkowski:1993fg,Caprini:2007xq,Huber:2008hg,Jinno:2016vai,Jinno:2017fby,Konstandin:2017sat,Cutting:2018tjt,Cutting:2020nla,Lewicki:2020jiv,Ellis:2020nnr}, the sound waves~\cite{Hindmarsh:2013xza,Giblin:2014qia,Hindmarsh:2015qta,Hindmarsh:2017gnf}, and the turbulence~\cite{Kamionkowski:1993fg,Caprini:2006jb,Caprini:2009yp,Kosowsky:2001xp,Gogoberidze:2007an,Niksa:2018ofa}. 
The parameter $\kappa_i$ are given by 
\begin{align}
 &\kappa_{\rm bubble} = \frac{\rho_{\rm bubble}}{\rho_{\rm vac}}, 
 \\
 &\kappa_{\rm SW} = \frac{\rho_{\rm SW}}{\rho_{\rm vac}}, 
 \\
 &\kappa_{\rm turb} = \frac{\rho_{\rm turb}}{\rho_{\rm vac}}, 
\end{align}
where 
$\rho_{\rm bubble}$, 
$\rho_{\rm SW}$, and $\rho_{\rm turb}$ are the energy densities of 
a thin shell around the bubble wall, 
bulk motion of the fluid, 
and turbulence, respectively. 
Here, $\rho_{\rm bubble}$ should be evaluated just before the end of the phase transition while $\rho_{\rm SW}$ and $\rho_{\rm turb}$ should be evaluated just after the phase transition.

As we stated in the main part of this paper, $\tilde{\Omega}_{\rm GW}$ is a function of $k/\beta$ and $v_w$ for the GW emission from the bubble collision. 
On the other hand, 
it includes a factor of $\beta/H_*$ 
and other time scales 
for the GW emission from the sound wave and magnetohydrodynamics turbulence 
because the duration of the GW emission from these sources is relatively long~\cite{Caprini:2009yp}. 
It is enhanced by a factor of $(\beta/H_*)(1-1/\sqrt{1+2 t_{\rm sw} H_*})$ 
for the GW emission from the sound wave, where 
$t_{\rm sw} \simeq (8 \pi)^{1/3} v_w / (\beta U_f)$ is the sound-wave period~\cite{Ellis:2018mja,Ellis:2019oqb,Ellis:2020awk,Guo:2020grp}
and $U_f^2 \simeq (3/4) \kappa_{\rm sw} \alpha/(1 + \alpha)$ is the root-mean-square four-velocity of the plasma~\cite{Hindmarsh:2015qta,Hindmarsh:2017gnf}. 
In our numerical calculation, we take  $U_f^2=\kappa_{\rm sw}$. 
On another hand, 
it is enhanced by a factor of $(\beta/H_*) (\rho_{\rm DR}/\rho_{\rm turb} )^{1/2}$ 
for the emission from the turbulence~\cite{Niksa:2018ofa}.

The GW spectra produced by the bubble collision~\cite{Huber:2008hg}, sound waves~\cite{Hindmarsh:2015qta,Hindmarsh:2020hop}, and  turbulence~\cite{Caprini:2009yp,Binetruy:2012ze}%
\footnote{
Since $\tilde{\Omega}_{\rm GW,turb}$ is the one for the magnetohydrodynamics turbulence, 
we implicitly assume that the dark sector has a gauge interaction. 
As discussed in Ref.~\cite{Hindmarsh:2015qta}, however, this contribution is subdominant 
and this assumption does not change our result.
}
are given by 
\begin{align}
&\tilde{\Omega}_{\rm GW,bubble}(f) \simeq
    1.0 \, \lmk \frac{0.11 v_w^3}{0.42 + v_w^2} \rmk 
    F_{\rm bubble} (f), \\
&\tilde{\Omega}_{\rm GW,sw}(f) \simeq
    0.16 \, v_w\left(\frac{\beta}{H_*}\right)
    \lmk 1-\frac{1}{\sqrt{1+2 t_{\rm sw} H_*}} \rmk
    F_{\rm sw} (f), \\
&\tilde{\Omega}_{\rm GW,turb}(f) \simeq
    20 \, v_w\left(\frac{\beta}{H_*}\right)
    \lmk 
    \frac{\kappa_{\rm turb} \alpha}{1+\alpha} \rmk^{-1/2}
    F_{\rm turb} (f), 
\end{align}
respectively, where 
\begin{align}
 &F_{\rm bubble} (f) = \left(\frac{3.8 \lmk f/f_{\rm bubble} \rmk^{2.8}}{1+2.8 \lmk f/f_{\rm bubble} \rmk^{3.8}}\right),
 \\
 &F_{\rm sw} (f) = \left(\frac{f}{f_{\rm sw}}\right)^3\left(\frac{7}{4+3\left(f/f_{\rm sw}\right)^2}\right)^{7/2},
 \\
 &F_{\rm turb} (f) = \frac{(f/f_{\rm turb})^3}{(1+(f/f_{\rm turb}))^{11/3}(1+8\pi f/h_*)}. 
\end{align}
Here, $f$ is the frequency today 
and $h_*$ is the Hubble parameter at $T = T_*$ that is redshifted today: 
\begin{align}
&h_*\simeq
   1.1\times 10^{-8}\,{\rm Hz}\times\left(\frac{T_*}{0.1\,{\rm GeV}}\right)\left(\frac{g_*}{10.75}\right)^{1/2} \left(\frac{g_{*s}}{10.75}\right)^{-1/3}\ .
\end{align}
The peak frequencies $f_{\rm bubble}$, $f_{\rm sw}$, and $f_{\rm turb}$ are given by
\begin{align}
&f_{\rm bubble}\simeq
   1.1 \times 10^{-8}\,{\rm Hz}
   \lmk \frac{0.62}{1.8 - 0.1 v_w + v_w^2} \rmk 
   G(\beta/H_*, T_*) \,,\\
  &f_{\rm sw}\simeq
   1.3\times 10^{-8}\,{\rm Hz}\times \frac{1}{v_w} 
   G(\beta/H_*, T_*) \,,\\
&f_{\rm turb}\simeq
   1.9\times 10^{-8}\,{\rm Hz}\times \frac{1}{v_w}
   G(\beta/H_*, T_*)\,,
\end{align}   
where 
\begin{align}
   G(\beta/H_*, T_*) = \left(\frac{\beta}{H_*}\right)\left(\frac{T_*}{0.1\,{\rm GeV}}\right)\left(\frac{g_*}{10.75}\right)^{1/2} \left(\frac{g_{*s}}{10.75}\right)^{-1/3}\ .
\end{align}

If the vacuum bubble interacts with the thermal plasma strongly enough and if the energy density of the thermal plasma in the dark sector is large enough, 
the accelerating bubble wall receives a large friction from the interaction with the thermal plasma~\cite{Bodeker:2009qy}. 
As a result, the bubble wall velocity may reach a terminal velocity to balance between the friction effect and the pressure due to the false-vacuum energy. 
Then most of the kinetic energy of the accelerating bubble wall is injected into the thermal bath. In this case, sound wave and turbulence of the plasma are the main sources of GWs. 
On the other hand, 
if the interaction between the vacuum bubble and the thermal plasma is not strong enough or the energy density of the thermal plasma in the dark sector is not large enough, 
the bulk energy is negligible and the bubble collision is the dominant source of GWs~\cite{Caprini:2015zlo}. This is the case for e.g., a significant supercooling 
(i.e., the case of $\Delta \gg 1$). 
In the intermediate case, 
all three sources are relevant.

\bibliography{bib}
\bibliographystyle{utphys}

\end{document}